\documentclass[rmp,aps,onecolumn]{revtex4}
\usepackage{graphicx}

\newcommand{\be}{\begin{equation}}
\newcommand{\ee}{\end{equation}}
\newcommand{\bea}{\begin{eqnarray}}
\newcommand{\eea}{\end{eqnarray}}
\newcommand{\ep}{\epsilon}

\newcommand{\zh}{{\bf e}_{||}}
\newcommand{\bF}{{\bf F}_g}
\newcommand{\bu}{{\bf u}}

\newcommand{\pd}[2]{\frac{\partial #1}{\partial #2}}
\newcommand{\etal}{{\it et al. }}
\newcommand{\ie}{{\it i.e. }}
\newcommand{\eg}{{\it e.g. }}

\newcommand{\oo}[1]{{\cal O}\left( #1 \right)}
\begin{document}

\title{A mathematical model for top-shelf vertigo: the role of sedimenting otoconia in BPPV}

\author{Todd M. Squires$^1$, Michael S. Weidman$^2$, Timothy C. Hain$^3$ and Howard A. Stone$^2$}
\affiliation{$^1$Departments of Applied and Computational Mathematics and Physics, California Institute of Technology, Pasadena, CA 91125\\
$^2$Division of Engineering and Applied Sciences, Harvard University, Cambridge, MA 02138\\
$^3$Northwestern University Medical School, Chicago, IL  60611}

(to be published in Journal of Biomechanics)

\begin{abstract}
Benign Paroxysmal Positional Vertigo (BPPV) is a mechanical disorder of the vestibular system in which calcite particles called otoconia interfere with the mechanical functioning of the fluid-filled semicircular canals normally used to sense rotation.  Using hydrodynamic models, we examine the two mechanisms proposed by the medical community for BPPV: cupulolithiasis, in which otoconia attach directly to the cupula (a sensory membrane), and canalithiasis, in which otoconia settle through the canals and exert a fluid pressure across the cupula.  We utilize known hydrodynamic calculations and make reasonable geometric and physical approximations to derive an expression for the transcupular pressure $\Delta P_c$ exerted by a settling solid particle in canalithiasis.  By tracking settling otoconia in a  two-dimensional model geometry, the cupular volume displacement and associated eye response (nystagmus) can be calculated quantitatively.  Several important features emerge:  1)  A pressure amplification occurs as otoconia enter a narrowing duct;  2) An average-sized otoconium requires approximately five seconds to settle through the wide ampulla, where $\Delta P_c$ is not amplified, which suggests a mechanism for the observed latency of BPPV; and  3)  An average-sized otoconium beginning below the center of the cupula can cause a volumetric cupular displacement on the order of $30$ pL, with nystagmus of order $2^\circ$/s, which is approximately the threshold for sensation.  Larger cupular volume displacement and nystagmus could result from larger and/or multiple otoconia.
\end{abstract}
\maketitle

\section{Introduction}
\label{sec:intro}

Benign Paroxysmal Positional Vertigo (BPPV) is the most commonly
diagnosed vertigo syndrome \cite{brandt91}, with a recent study suggesting that it 
affects 9\% of older persons \cite{oghalai00}.  BPPV is characterized by sudden attacks of dizziness and nausea
triggered by changes in head orientation, and specifically afflicts the posterior canal.  The disorder has earned the common name ``top-shelf vertigo'', since attacks often occur when the head is suddenly tilted back, such as when looking at objects on the top shelf.  Other clinical features of BPPV include a 5-10 second {\sl latency} between the head tilt and the onset of vertigo, and a {\sl fatiguable} response which lessens with repeated head maneuvers \cite{brandt93}.  Although the condition is not life-threatening, the disorientation brought on by attacks is severely discomforting, and can cause nausea, accidents and injuries.

BPPV is caused by a mechanical dysfunction of the vestibular system in the inner ear, whose fluid-filled {\sl semicircular canals} normally act to detect rotation via deflections of the sensory membraneous {\sl cupula}.  In BPPV, calcite particles (otoconia) are believed to interfere with the normal operation of the semicircular canals, falsely inducing a spinning sensation when in fact no rotary motion of the head is actually occurring.

Two primary theories for the cause of BPPV have been advanced by the medical community:
{\sl cupulolithiasis}, in which otoconia are directly attached to the cupula, and {\sl canalithiasis}, 
in which otoconia freely sediment through the canals and exert a fluid pressure on the cupula.  A 
consensus is emerging that canalithiasis is the more likely mechanism for BPPV, supported in part by the
largely successful clinical technique used in BPPV treatment:  therapeutic head maneuvers \cite{epley92,semont88,brandt80} designed to drive otoconia all the way around and out of the canal, so that they settle in the vestibule/utricle.  Further support for canalithiasis is reviewed in the work of 
Brandt and Steddin \cite{brandt93}, who compare the two mechanisms and conclude that canalithiasis is better able to explain the latency and fatiguability of the disorder.  Recently, however, Buckingham \cite{buckingham99} questioned the canalithiasis interpretation for the maneuvers, suggesting that settling particles should exert a transcupular pressure from the vestibule as well as from the narrow duct.  Finally, House and Honrubia \cite{house03} cite various clinical observations, some of which are consistent with canalithiasis, and some with cupulolithiasis, and suggest that in fact both mechanisms are viable and occur.

Despite extensive quantitative modelling of normal vestibular functioning, the description and debate on mechanisms for BPPV have been purely qualitative.  Without a quantitative analysis of these mechanisms, it is impossible to know whether either mechanism is physically capable of producing a response of the magnitude experienced in BPPV.  Recently, House and Honrubia \cite{house03} have taken an important first step toward a quantitative mathematical model of canalithiasis and cupulolithiasis.  Specifically, they postulate that a settling sphere exerts a transcupular pressure 
$\Delta P_H = \bF\cdot\zh/3\pi b(s_p)^2,$ where $\bF$ is the force on the particle due to gravity (corrected for buoyancy), $\zh$ is the unit vector oriented along the canal axis, $b(s)$ is the canal radius (which varies with centerline coordinate $s$), and $s_p$ is the  particle location.  This formula was argued based upon the following assumptions: (i) the component of the force perpendicular to the channel exerts no transcupular pressure; (ii) the fluid pressure is uniform across the canal cross-section containing the otoconium; (iii) the applied transcupular pressure arises {\sl only} from the pressure-driven component of the sphere's drag, and not from viscous shear stress (whence the factor of 1/3); and (iv) (not explicitly stated) the Stokes settling velocity and pressure drag for a sphere are unaffected by canal walls.  

The present work builds on and extends the results of House and Honrubia in several ways.  First, starting with the  Stokes equations for viscous flow, we derive the transcupular pressure $\Delta P_c$ resulting from a small particle settling through a fluid-filled channel.  Significantly, both qualitative and quantitative corrections to $\Delta P_H$ exist, since assumptions (ii-iv) are not supported by solutions to the Stokes equations.

Our expression for $\Delta P_c$ follows directly from the equations of hydrodynamics, is independent of particle shape or even size (so long as the particle is small compared to the channel), and is valid for a canal of arbitrary (but slowly-varying) geometry.  Furthermore, we consider two-dimensional motion of the otoconia (\ie head maneuvers which are in the canal plane), so that otoconia settle away from the canal centerline.  

We simulate BPPV episodes by tracking the cupular volume displacement due to the transcupular pressure exerted by the settling otoconia.  The cupular volume displacement is frequently assumed to be proportional to the `sensed' rotation rate.  Under such an assumption, the  `perceived' rotation rate (or, equivalently, the measured nystagmus) in our study can be obtained directly from the simulated cupular volume displacement.  There is evidence, however, that additional neural processing (called {\sl velocity storage}) occurs in the brain stem, and that the (measurable) nystagmus response reflects this process \cite{robinson77,raphan79}.  For completeness, we also simulate the nystagmus response using a `velocity storage' model of neural processing.  We emphasize, however, that our fundamental picture and analysis of canalithiasis and cupulolithiasis do {\sl not} depend on velocity storage.  We compare our model of canalithiasis to a model of cupulolithiasis, and demonstrate that canalithiasis can actually give a {\sl stronger} nystagmus response, with or without velocity storage.  This agrees with the conclusions of House and Honrubia.

\begin{figure}
\begin{center}
\includegraphics[height=3in]{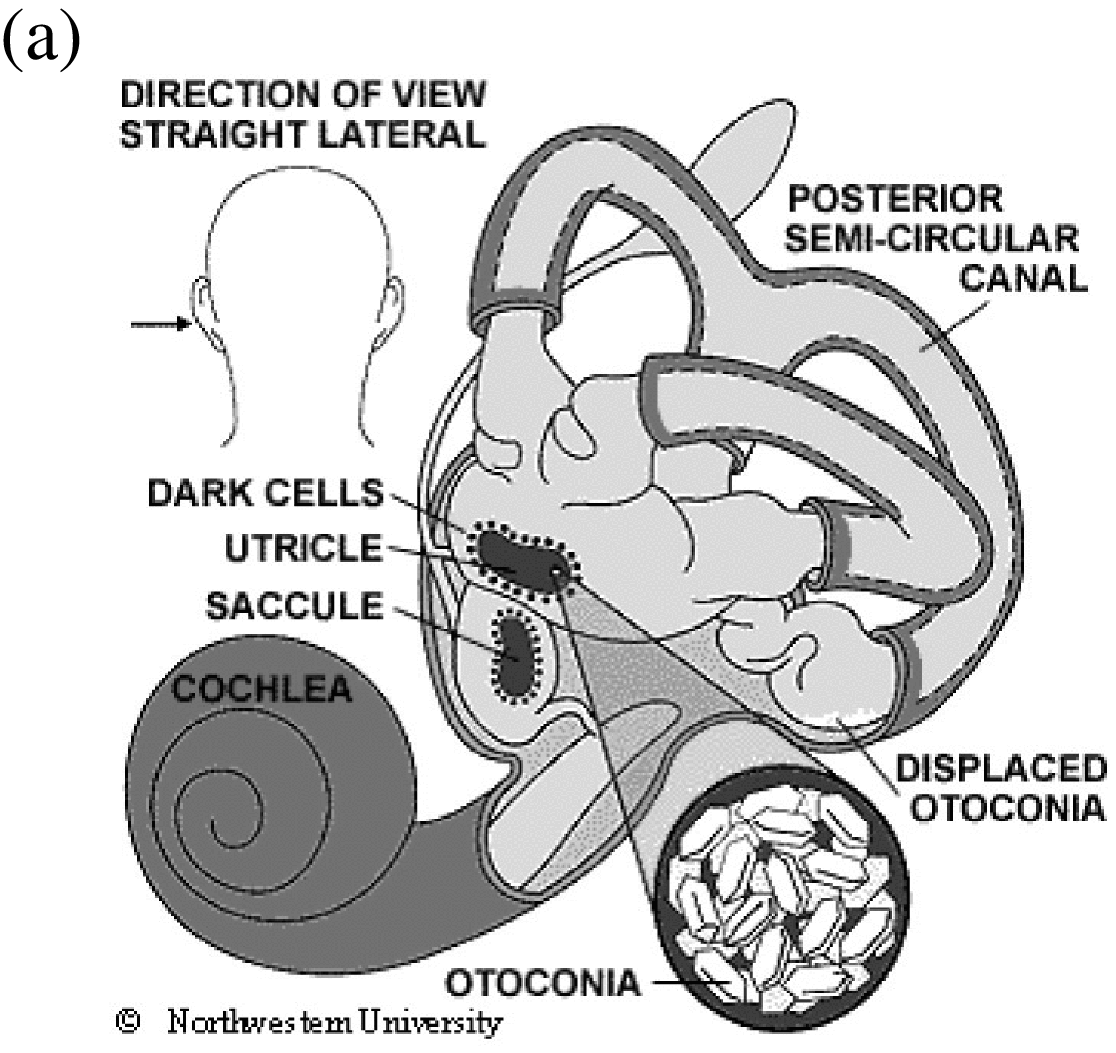}\hfill
\includegraphics[height=3in]{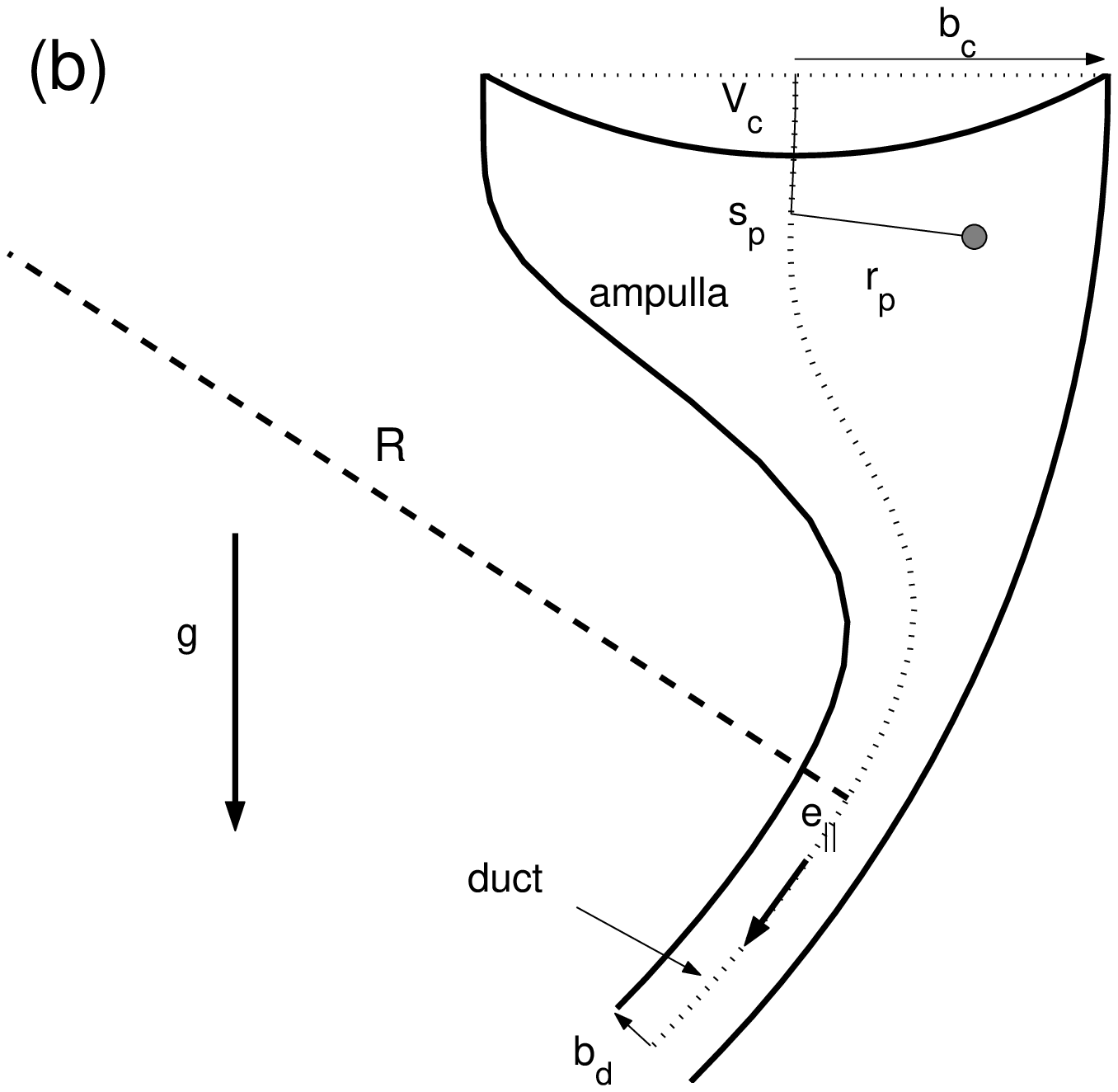}
\caption{\label{fig:defscc} a)  The human vestibular system consists of the semicircular canals for sensing rotation, the otolithic organs (saccule and utricle) for sensing linear/gravitational acceleration, and the cochlea for auditory sensing.  Displaced otoconia, perhaps from the otolithic organs, are believed to collect in the posterior canal and cause BPPV.  b)  Schematic of the model geometry for the semicircular canals (defined in Eq. (\ref{eq:defscctwo})), which is chosen to resemble measurements \cite{curthoys87}.  A torus of major radius $R$ consists of a thin circular duct of radius $b_d$, and a thicker region consisting of the ampulla and the vestibule (located above the cupula, not shown here). The ampulla is spanned and separated by the membraneous cupula of radius $b_c$, and the vector $\zh$ points along the (local) canal centerline, from the ampullar side of the cupula towards the vestibular side.  The position of a sedimenting otoconium is specified by its distance $s_p$ along the centerline of the canal and its distance $r_p$ from the centerline of the canal.  The translating otoconium exerts a pressure field that displaces the cupula by a volume $V_c$.}
\end{center}
\end{figure}

\section{Methods}

In this section, we derive a simplified model that includes the physics of settling otoconia and the processing of a velocity storage mechanism in order to capture the essential characteristics of canalithiasis, without requiring a detailed numerical investigation.  Five components required for model simulations will be developed:  (i) the basic functioning of the semicircular canals; (ii) an equation for the volumetric cupular displacement $V_c$ to an externally-applied transcupular pressure difference $\Delta P_c$; (iii) the fluid pressure field set up by a settling particle of a given size, density, and position, giving the pressure difference $\Delta P_c$ that distorts the cupula;  (iv) the particle's settling velocity, accounting for Stokes drag, hydrodynamic interactions with channel walls, and background endolymph flow; and (v) the velocity storage mechanism.  

\subsection{Overview of normal vestibular functioning}
\label{sec:models}

The human vestibular system consists of two sets of organs:  the otolithic organs, which sense linear acceleration and gravity, and the semicircular canals, which sense angular motion.  Each ear has three semicircular canals that are oriented in a mutually orthogonal fashion in order to sense rotations about each of three axes (Fig. \ref{fig:defscc}a).  The canals are filled with endolymph of density $\rho$ and viscosity $\mu$ similar to that of water.
Each canal consists of a narrow {\sl duct} connected to thicker {\sl ampulla} and {\sl vestibule} regions, which are in turn separated by the membraneous cupula.

When the canal experiences an angular acceleration, inertia causes a lag in endolymph motion, distorting the cupula by a volume $V_c$. The cupular distortion triggers activation of embedded sensory hair cells, signalling the angular acceleration/velocity to the brain and causing a compensatory eye motion called {\sl nystagmus}.   Under sustained rotation, the elastic restoring force of the cupula pushes endolymph back through the canal, and the cupula slowly relaxes to its undisplaced position.

Neural signal processing in the brain stem modifies the transmitted signal in a process called velocity storage \cite{robinson77,raphan79}, which lengthens the typical time constant of the associated sensation and nystagmus.  The typical time constant for cupular relaxation ($\tau_c \approx 4$ s) is transformed into a nystagmus time constant of about $\tau_v \approx 21$ s for rotation about the vertical axis of the head (yaw), where the lateral semicircular canal is stimulated.  Velocity storage may be less effective in the posterior canal, or might also be less effective for an unnatural vestibular stimulation such as BPPV.  Nevertheless, since experimental data regarding velocity storage in the posterior canal is presently unavailable, we assume it to be the same as in the lateral canal.

The model geometry (Fig. \ref{fig:defscc}b) we use in our simulations is chosen to closely resemble a semicircular canal --- a two-part torus of major radius $R$, with a thick ampulla region and a thin duct region.  We represent the inner and outer radii of the canal as functions of angle $(0<\theta\le 2 \pi)$, 
\be
R_{\rm out} = R + b_d\,\,\,\,\mbox{and} \,\,\,\,R_{\rm in}(\theta)  =  R-\left[A+B \tanh\left(\frac{\theta-\theta_1}{\delta}\right)\right]\label{eq:defscctwo},
\ee
where $A =2b_c+ b_d$ and $B = 2b_c-b_d$.  The canal narrows from radius $b_c$ in the ampulla to $b_d$ in the duct roughly over an angle $2\delta$, centered around $\theta_1$. We define the cupula to be located at $\theta=0$, and choose numerical values (Table \ref{tab:parameters}) to reflect the measurements of Curthoys and Oman \cite{curthoys87}, with channel radii chosen to have cross-sectional areas equal to the roughly elliptical channels measured.  We use $\beta_d R$ and $\beta_u R$ (not shown) for the lengths of the duct and vestibule/utricle to be consistent with the geometry of Van Buskirk \etal \cite{vanbuskirk76}.  The model geometry, though simplified, captures the most important shape variations from a hydrodynamic standpoint. 

\begin{table}
\begin{center}
\begin{tabular}{|c|l|c|c|} 
\hline 
\multicolumn{4}{|c|}{Physical parameters for human semicircular canals} \\ 
\hline
$R$ 				& Major radius of semicircular canal 	& 3.2 mm 						& \cite{curthoys87}\\
$b_d$ 			& Duct radius 												& 0.16 mm 					& ''\\
$b_c$ 			& Mean radius of cupular partition		& 0.68 mm 					& ''\\
$\theta_1$	& Angle where channel tapers					& 36 $^\circ$				& ''\\
$\delta$		&Angle over which channel tapers 			& 5.4$^\circ$						& ''\\
$\beta_d$ 	& Angle subtended by duct 						& 250$^\circ$ 					& \cite{vanbuskirk76}\\
$\beta_u$ 	& Angle subtended by vestibule/utricle 	& 75$^\circ$  				& ''\\
$\rho$ 			& Endolymph density 									& 1.0 g/cm$^3$ 			& \cite{biomed}\\
$\mu$ 		& Endolymph viscosity 							  	& 0.01 g/cm/s			  & ''\\
$\rho_o$ 		& Otoconium density 									& 2.7 g/cm$^3$ 			& ''\\
$a$ 				& Otoconium radius 					 					& 0.5--15$\mu$m, 7.5 $\mu$m avg.				&\cite{campos90} \\
$\tau_c$		&Cupular time constant								&4.2 s							&\cite{dai99}\\
$\tau_v$		&Nystagmus time constant							&21 s								&\cite{malcolm68}\\
$K$					&Cupular elastic constant							& 13 GPa/m$^3$ 			& Calculated using $\tau_c$  in Eq. (\ref{eq:vanbuskirk})\\
$\lambda_1$	& First zero of Bessel Function $J_0$	& 2.4							 &\cite{vanbuskirk76}\\
$\Omega_{\rm th}$&Threshold rotation for sensation	&2$^\circ$/s &\cite{oman87}\\
${\bf g}$		& Gravitational acceleration 					& 9.8 m/s$^2$			&\\
\hline
\end{tabular}
\caption{\label{tab:parameters}Physical parameters and derived quantities for BPPV.}
\end{center}
\end{table}


\subsection{Mathematical model for endolymph flow in the semicircular canals}
\label{sec:physmod}

The classic mathematical model of the normal functioning of the semicircular canals, originally developed by Steinhausen \cite{steinhausen33}, treats endolymph displacement with an overdamped pendulum equation.  Two time scales appear in Steinhausen's equation:  a rapid time scale $\tau_f$, during which fluid moves ballistically through the canal, and a slow time scale $\tau_c$ over which the cupula relaxes back to its undistorted position.  
Subsequent theoretical work has typically worked towards calculating these time scales from the physical parameters and length scales characteristic of the vestibular system.  Examples include the transient and quasi-steady fluid motion in a torus with thin (duct) and thick (vestibule/utricle) sections \cite{vanbuskirk76}, ducts with non-constant and non-circular cross section \cite{oman87}, allowances for high-accelerations by calculating the fluid flow near the cupula partition rather than assuming a Poiseuille profile \cite{rabbitt96}, fluid flow through a possibly porous cupula \cite{damiano99}, and the effects of multiple hydrodynamically-connected canals \cite{muller88}.
Because the approach of Van Buskirk \etal gives results that are largely consistent with these revised models in
the low frequency and velocity regimes considered here, we use a similarly simplified geometry in the present article.  It should be straightforward to extend the present work to a more complicated (but realistic) three-dimensional model of the semicircular canals.

Van Buskirk \etal explicitly calculated the time constants in terms of the system dimensions and parameters as
\be
\tau_c = \frac{8 \mu \beta_d R}{\pi b_d^4K }\,\,\,\,\mbox{and}\,\,\,\, \tau_f = \frac{\rho b_d^2}{\lambda_1^2 \mu}\label{eq:vanbuskirk},
\ee
where $\lambda_1$ is the first zero of the Bessel function $J_0(\lambda),$ \ie $\lambda_1 \approx 2.4$.  All physical parameters are known other than the elastic constant $K$ of the cupula, here chosen to give the measured value ($\tau_c = 4.2$ s) for the human cupular time constant \cite{dai99}.  Following a step increase in angular velocity $\alpha(t) = \Omega_0 \delta(t)$, the endolymph in the duct (and therefore the cupula) experiences a maximum displacement of volume
\be
V_c =\frac{4\rho\Omega_0(1+\beta_u/\beta_d)\pi R b_d^4}{\lambda_1^4 \mu},
\label{eq:maxdisp}
\ee
after a time $\tau_f$, after which the cupula relaxes back on a time scale $\tau_c$.  Using the values in Table \ref{tab:parameters}, the cupular volume displacement is related to rotation rate via $\Omega_0/V_c\approx 5.6 \times 10^{-2}\,^\circ$/pL s, and a threshold rotation rate of $2^\circ$/s corresponds to a cupular volume displacement of 35 pL.

The influence of fluid inertia occurs over millisecond time scales, whereas the state of the fluid, cupula, and otoconium all change on significantly longer time scales.  We therefore neglect fluid inertia throughout this study.  With this approximation, the cupular membrane obeys a simple balance of pressures, in which three terms are important.  A cupular membrane that is distorted to displace a volume $V_c$ exerts an elastic restoring force that contributes an amount $-K V_c$ to the transcupular pressure difference.  Second, by conservation of mass, instantaneous cupular and endolymph volume displacements must be equal.  The resulting viscous force contributes an amount $-\gamma \dot{V_c}$ to the transcupular pressure, where the viscous coefficient $\gamma$ will be calculated shortly.  (Throughout this paper, dots denote time derivatives.)
The last term to enter the pressure balance is an externally applied, time-dependent pressure difference $\Delta P_c(t)$, which in our case is given by the fluid pressure set up by a settling otoconium.  
These three pressures balance, giving an equation for cupular volume displacement
\be
-K V_c - \gamma \dot{V_c} +\Delta P_c(t)=0.
\label{eq:eom}
\ee
In normal vestibular functioning, $\Delta P_c$ arises from angular acceleration of the canals, in which case Eq. (\ref{eq:eom}) is consistent with Steinhausen's model \cite{steinhausen33} if the effect of fluid inertia is neglected.

The viscous resistance coefficient $\gamma$ for quasi-steady fluid flow through the duct can be derived in a 
straightforward manner.  Fluid in a straight channel of length $\beta_d R$ and circular cross-section of radius $b_d$,
subject to an applied pressure difference $\Delta P$, moves with a parabolic velocity profile directed along the cylinder axis $\zh$ \cite{leal92}, 
\be
\bu(r) = \frac{\Delta P}{4 \mu \beta_d R}\left(b_d^2-r^2\right)\zh.
\label{eq:poiseuille}
\ee
Note that we have assumed the usual no-slip condition for a solid/fluid interface.
From the flow rate $\dot{V}_c =\pi b_d^4\Delta P/8 \mu \beta_d R$, the viscous coefficient $\gamma$ is shown to be
\be
\gamma = \frac{8 \mu \beta_d R}{\pi b^4}.
\label{eq:viscres}
\ee 

This result can be expected to provide a very good approximation for the low-Reynolds-number flow 
in the slightly curved canal duct and slowly-varying geometries of interest here.  The pressure drop along the duct (which varies approximately with $\approx b_d^4/\beta_d$) is approximately a thousand times greater than that along the vestibule/utricle and ampulla (which varies with $b_c^4/\beta_u$).  Therefore, we approximate the viscous resistance in the canal as occuring in the duct alone.  Furthermore, we approximate the channel as locally straight, because the radius of curvature of the channel $R$ is large compared to the duct radius.  This introduces errors of magnitude $\oo{b_d/R}\approx 0.05$ in the axial flow and $\oo{b_d^2/R^2}\approx 10^{-3}$ in the flow rate \cite{leal92}.  Finally, although the channel radius is not constant, it typically varies slowly within the duct.  

The cupular volume displacement $V_c$ that results from an applied pressure $\Delta P_c(t)$ 
is given by the solution of Eq. (\ref{eq:eom}),
\be
V_c(t) = \frac{1}{\gamma}\int_{-\infty}^t \Delta P_c(t')e^{-(t-t')/\tau_c}dt',
\label{eq:xcsol}
\ee
where $\tau_c$ is given in Eq. (\ref{eq:vanbuskirk}).

\subsection{Pressure drop exerted by sedimenting otoconia}
\label{sec:pressuredropfromsed}

The time-dependent pressure $\Delta P_c(t)$ in Eq. (\ref{eq:xcsol}) is provided by the stress field set up by a settling otoconium, which we model as a sphere of radius $a\ll b_d$ and density $\rho_o$ settling through
the fluid due to gravity ${\bf g}$.  In Stokes flows, the velocity and pressure fields around a translating particle are insensitive to the detailed shape or size of the particle outside of the immediate vicinity of the particle.  Rather, they depend only on the total force exerted by the particle.  Therefore, we can approximate the fluid flow due to the particle as that of an equivalent `point force,' since $a \lesssim b_d/15$ even in the narrowest part of the curved duct.

In Appendix \ref{app:recip}, we demonstrate that the difference in pressure between the fluid in front and behind a small particle settling in a (possibly curved) circular cylinder of local radius $b(s)$ 
is given by
\be
\Delta P_c = 2 \frac{\bF \cdot \zh}{\pi b(s_p)^2}\left(1-\frac{r_p^2}{b(s_p)^2}\right).
\label{eq:pressuredrop}
\ee
Here $\zh$ and $s_p$ are defined as in $\Delta P_H$, and $r_p$ measures the radial distance of the particle from
the canal centerline (Fig. \ref{fig:defscc}b).  The gravitational force on the particle is given by
\be
{\bF} = \frac{4}{3} \pi a^3 \Delta \rho {\bf g},
\label{eq:buoyantforce}
\ee
where $\Delta \rho = \rho_o-\rho$ is the density difference between the otoconium and the endolymph.
Eq. (\ref{eq:pressuredrop}) was originally derived for a straight circular cylinder \cite{brenner58}, but the physics of low-Reynolds-number flows allows it to be applied, as a first approximation, for channels whose geometry (\eg  radius or orientation) varies slowly.  Due to the closed geometry of the semicircular canals, the pressure difference $\Delta P_c$ set up by the flow due to the settling sphere is exerted across the cupula, so that the forcing $\Delta P_c(t)$ in Eq. (\ref{eq:xcsol}) is given by Eqs. (\ref{eq:pressuredrop}) and (\ref{eq:buoyantforce}).  

The radial and axial positions $(r_p,s_p)$ of the otoconium determine the pressure $\Delta P_c$
exerted on the cupula via Eq. (\ref{eq:pressuredrop}), to which the cupula responds via Eq. (\ref{eq:xcsol}). 
Particle motion affects the pressure exerted on the cupula in three ways:  (i)  When otoconia fall from the ampulla (radius $b(s_p)=b_c$) into the narrow duct (radius $b(s_p)=b_d)$, the transcupular pressure is amplified by a factor $b_c^2/b_d^2 \approx 18$; (ii) due to the duct's curvature, otoconia fall away from the centerline towards the walls ($r_p$ approaches $b(s_p)$), which reduces the transcupular pressure due to the factor $1-r_p^2/b(s_p)^2$ in Eq. (\ref{eq:pressuredrop}); and (iii) the unit vector $\zh$ changes direction as an otoconium falls, which changes the force projection ${\bf g} \cdot \zh$.  

Eq. (\ref{eq:pressuredrop}) is expected to hold with errors of order $(a/b(s_p))^2$ and $b(s_p)/R$ (up to roughly five percent), with three exceptions.  First, the `point force' approximation breaks down for particles located within a few particle radii of canal walls, which requires a more detailed analysis.  The pressure is then expected to be small, but not zero as predicted by Eq. (\ref{eq:pressuredrop}).  Second, the analysis leading to Eq. (\ref{eq:pressuredrop}) does not hold for particles near the cupula.  In this case, the pressure distribution along the cross-sectional area of the cupula is not constant, and the total force on the cupula  (\ie pressure integrated over the cupular area) varies continuously from $\bF \cdot \zh$ (for a particle just below the cupula) to Eq. (\ref{eq:pressuredrop}) for a particle beyond a few channel radii from the cupula.  We have not accounted for these variations in our simulations for small particles (Sec. \ref{sec:bppvattacks}) because most of the cupular volume displacement occurs when Eq. (\ref{eq:pressuredrop}) is valid.  Third, if the particle occupies a significant portion of the canal (\eg ``canalith jam'' \cite{epley95}) the transcupular pressure would be greater than given in Eq. (\ref{eq:pressuredrop}).  

\subsection{Sedimentation velocity of otoconia}
\label{sec:settlingvelocity}
Here we give an approximate calculation of the sedimentation velocity of an otoconium, considering three possible corrections to the standard Stokes settling velocity: (i) particle inertia, (ii) advection with the background flow, and (iii) hydrodynamic interactions with the walls.  The first two corrections are shown to be negligibly small, whereas hydrodynamic interactions are small except when otoconia are very close to channel walls.  

Since the particle is small compared to the size of the channel, a good first approximation to its settling speed is given simply by Stokes' formula \cite{leal92}
\be
U_s = \frac{F}{6 \pi \mu a} = \frac{2 \Delta\rho a^2g}{9 \mu}\approx 0.2\mbox{mm/s},
\label{eq:stokesvel}
\ee
using values from Table \ref{tab:parameters}. 

The influence of particle inertia can be estimated by balancing the inertial term $m \ddot{x}$, where $x(t)$ denotes position, against viscous resistance $6\pi \mu a \dot{x}$, from which an inertial time scale emerges
\be
\tau_i \approx \frac{m}{6\pi\mu a} \approx \frac{2\rho a^2}{9 \mu} \approx 10^{-2}\, {\rm ms},  
\ee
during which time a settling otoconium moves only a few nanometers.  Thus inertia is negligibly small for settling otoconia, and is neglected here.  Note that inertial effects such as the Magnus lift (on rotating particles) are likewise negligibly small.

Next, we consider the influence of the background velocity of the endolymph, which is set into motion by the displaced cupula.  A cupula displacing a volume per unit time $\dot{V}_c$ sets up a parabolic flow in the duct (Eq. (\ref{eq:poiseuille})), whose maximum velocity is twice its average, or $2 \dot{V}_c/\pi b_d^2.$
From simulations (\eg Fig. \ref{fig:toroidbppv}), we estimate a cupular volumetric displacement rate of order $\dot{V}_c\lesssim$ 10 pL/s, giving a maximum endolymph velocity in the duct of order
$0.3 \mu$m/s --- approximately a thousand times smaller than the Stokes settling velocity, given by Eq. (\ref{eq:stokesvel}).  We thus neglect endolymph flow in computing particle motion.

Finally, we incorporate previously calculated hydrodynamic interactions between the particle and the walls of an enclosing cylinder.  Details can be found in appendix \ref{app:hydrocoupling}.  We list here the most significant features:  1)  Hydrodynamic interactions are largely insignificant away from channel walls, where the relative correction to Eq. (\ref{eq:stokesvel})  is of order $\oo{a/b} < .08$ unless the particle is very close to the wall \cite{hirschfeld84}.  
2)  Our approximations do not hold for otoconia very close to the wall, since non-spherical otoconia shapes and membraneous duct coatings \cite{curthoys87} may play significant roles.  Without reliable information about otoconium-wall interactions, we simply assume the otoconium to behave like a sphere, with near-wall parallel ($U_\|$) and perpendicular ($U_\perp$) velocities \cite{goldman67,oneill67,cooley68,happel,batchelor}
\be
U_\| \approx \frac{U_s}{\log(d/a)}\,\,\,{\rm and}\,\,\,U_\perp \approx U_s \frac{d}{a},
\label{eq:upara}
\ee 
where $d$ is the particle-wall gap.  We introduce a cutoff $d_c = 1$ $\mu$m for the particle-wall gap so that otoconia slide along the duct walls, as is required for the therapeutic maneuvers to be geometrically possible.  Likewise, the transcupular pressure exerted by a particle very close to the wall is not known; however, we expect it to be small.   

\subsection{Velocity storage as a model of neural processing}
\label{sec:vsm}

Our treatment thus far has concerned only the mechanical response of the system.  Signal processing is believed
to occur in the brain stem in a process called velocity storage \cite{robinson77,raphan79}.  While velocity storage
does not affect the mechanical response of the semicircular canals, it is believed to affect the (measurable) nystagmus response, denoted $\dot{E}$, and we therefore include it in the present work.  (Another modification involves hair cell and afferent {\sl adaptation}, which we do not treat due to its much longer (60-80 s) time scales).  The effect of velocity storage is to lengthen the long time constant of the horizontal semicircular canal $\tau_c \approx 4.2$ seconds \cite{dai99} to a longer time scale $\tau_v\approx 16-21$ seconds \cite{hain92,dai99}, which improves the performance of the canals for angular rate sensing, extending their bandwidth to lower frequencies.
The velocity storage mechanism is typically described in Laplace transform space
(indicated by tildes) using a transfer function \cite{robinson77, raphan79}
\be
\frac{\tilde{\dot{E}}(s)}{\tilde{\Omega}(s)} =\frac{1 + s \tau_c}{1 + s \tau_v}.
\label{eq:vor}
\ee
where $\Omega$ is the angular velocity transduced by the canal (Eq. (\ref{eq:maxdisp})).
Combining Eq. (\ref{eq:vor}) with the Laplace transform of Eq. (\ref{eq:eom}), we arrive at
an expression (in the time domain) for nystagmus as a function of transcupular pressure,
\be
\dot{E}(t) = \frac{\lambda_1^4}{32 \rho (\beta_u + \beta_d)R^2}\int_{-\infty}^t \Delta P_c(t')e^{-(t-t')/\tau_v}dt'.
\label{eq:nystagmussol}
\ee
This expression allows us to translate the mechanical effect of a sedimenting particle into
an equivalent sensation of motion, which is directly measurable as an eye movement.  Again, however,
we emphasize that the main affect of velocity storage is to lengthen the cupular relaxation time 
scale.  The basic qualitative and quantitative conclusions we draw from our simulations are essentially unaffected by whether or not velocity storage is incorporated, other than the time constant for nystagmus decay.

\section{Results}
\subsection{Canalithiasis}
\label{sec:bppvattacks}

In this section, we describe simulations of canalithiasis in which we numerically integrate the motion of an otoconium settling under gravity.  The orientation of the canal is held constant in these simulations, and the initial position of the otoconia is taken to be just below the cupula, since this represents the lowest point when the head is held upright.  The transcupular pressure exerted as the otoconium falls (Eq. (\ref{eq:pressuredrop})) is calculated and used to determine the volumetric cupular displacement $V_c$  (Eq. (\ref{eq:xcsol})) and nystagmus (Eq. (\ref{eq:nystagmussol})).  
We present here the results of a simulation of an average-sized otoconium starting just below the cupula in a model canal whose geometry (Fig. \ref{fig:defscc}b) closely resembles the semicircular canals.  In addition, in appendix \ref{sec:bppvstraight} we simulate a particle falling in a simpler geometry -- a straight, tapering channel -- to clearly illustrate the pressure amplification that occurs as the particle enters a narrowing canal.  

\begin{figure}
\begin{center}
\includegraphics[height=2.2in]{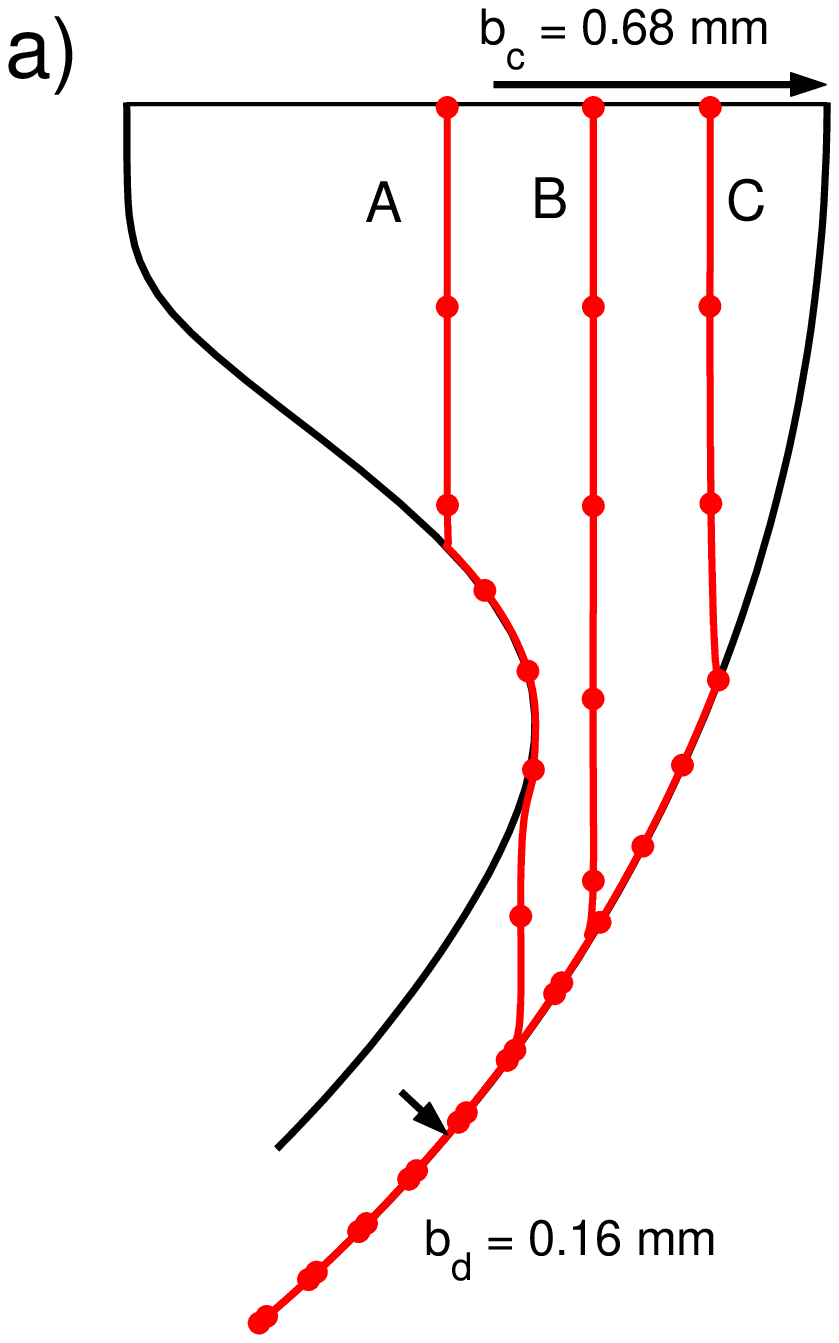}
\hfill\includegraphics[height=2.2in]{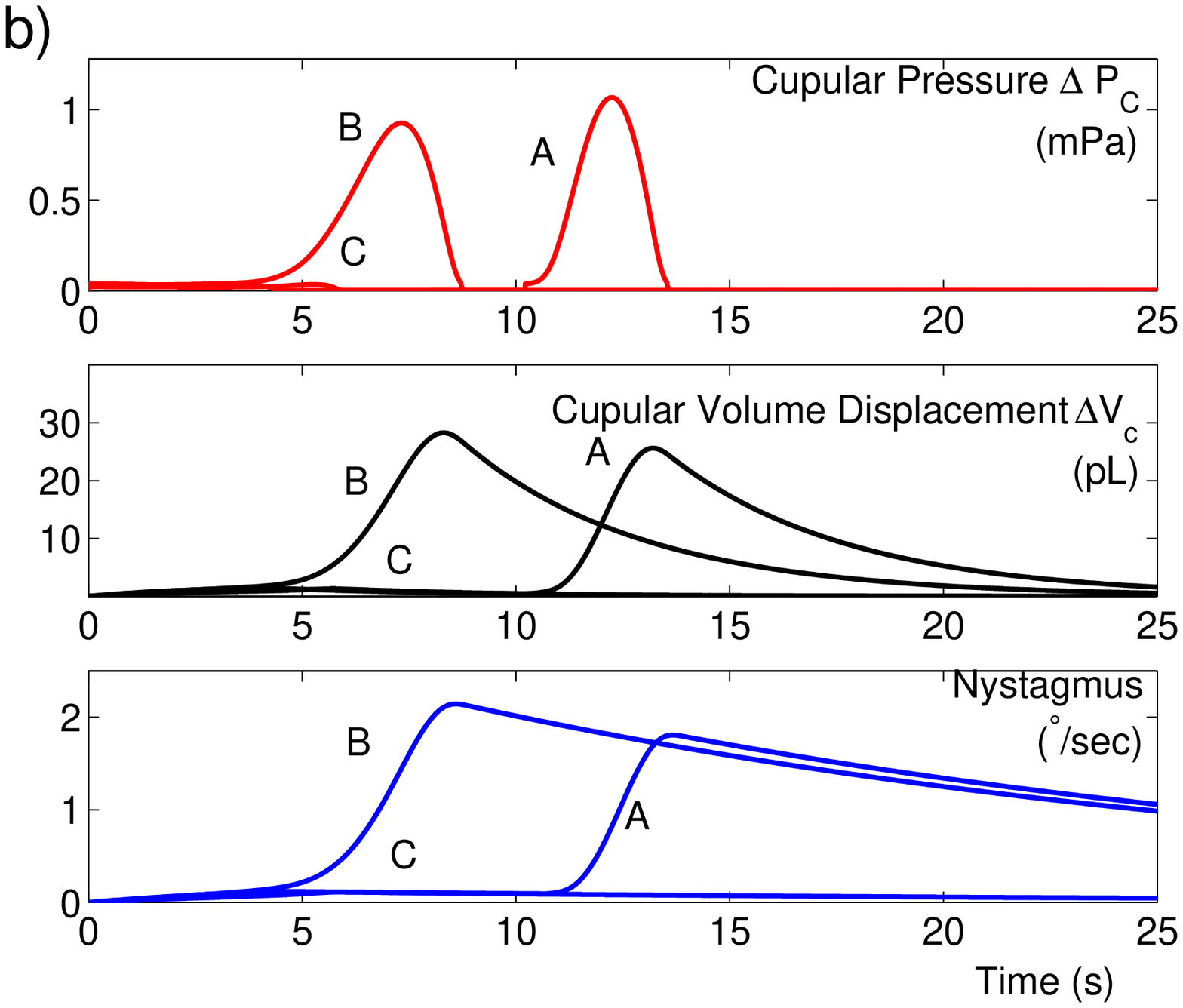}
\includegraphics[height=2.2in]{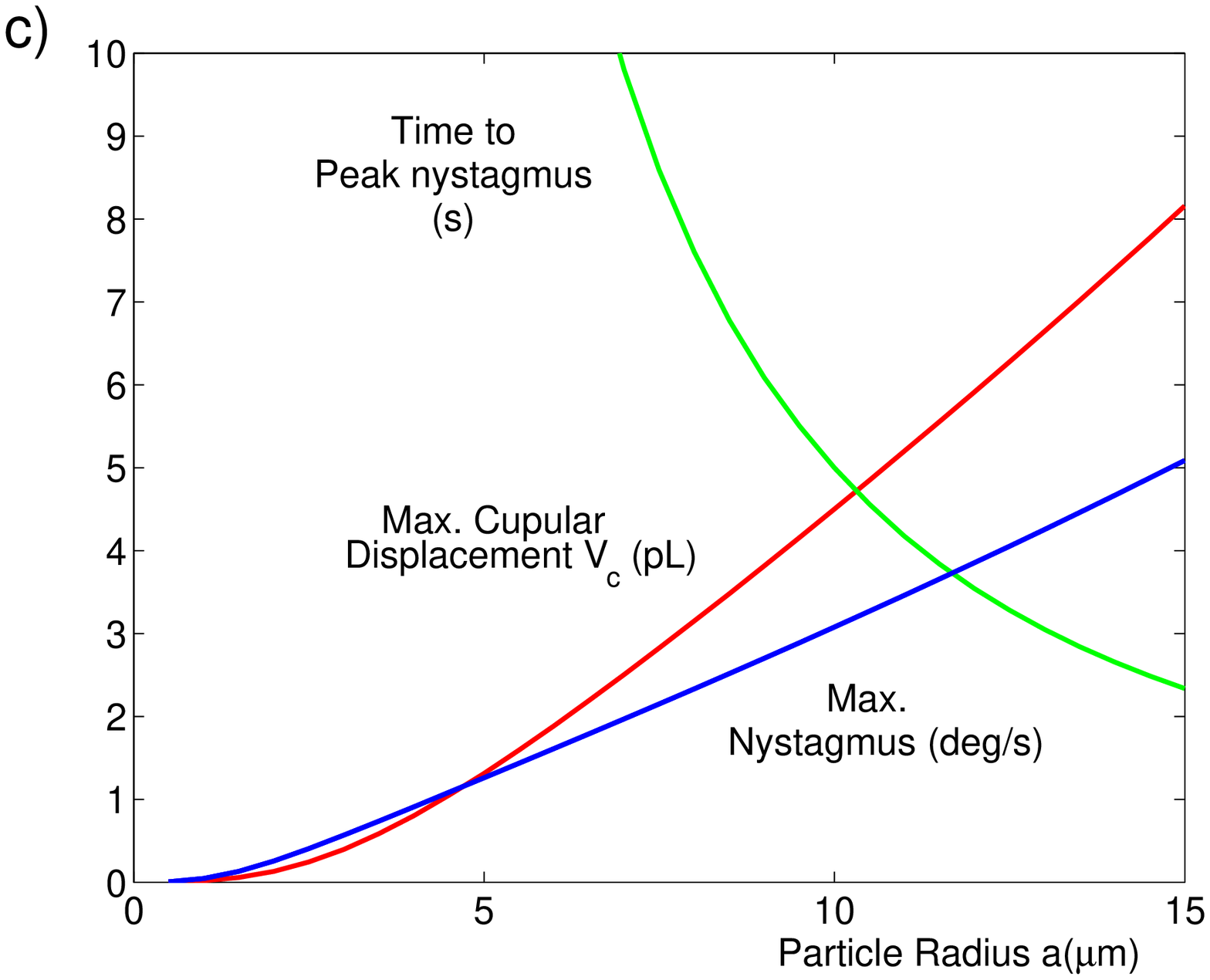}
\caption{\label{fig:toroidbppv} Simulated canalithiasis episodes occuring when otoconia, starting at different radial positions just below the cupula, fall through a tapering toroidal channel.  a) Otoconia trajectories, with dots to indicate the location of each otoconium at two-second intervals.  b) Results for (i) the pressure exerted on the cupula for the three trajectories, (ii) the cupular volume displacement for each of the three trajectories, and (iii) the nystagmus response, reflecting the sensed rotation velocity.  Note that Eq. (\ref{eq:maxdisp}) can be used to convert $\Delta V_c$ into nystagmus which gives peak values ($\dot{E}^{{\rm max}}\approx 2^\circ$/s) that are approximately the same as those obtained with velocity storage.  The largest cupular volume displacement occurs when the otoconium enters the narrow part of the canal, hydrodynamically amplifying the pressure it exerts by a factor of up to 36 times the naive estimate of $F_g/\pi b_c^2$ (which would occur during cupulolithiasis).  This occurs with a 6-13 second 
latency for trajectories $A$ and $B$, and never occurs for trajectory $C$, which hits the wall before entering the duct. c)  Simulations for particles of different radii, all started on trajectory B in the model semicircular canal geometry (Fig. \ref{fig:toroidbppv}a).  Peak nystagmus and cupular volume displacement increase roughly linearly with particle size $a$ for particles larger than a 3-5 microns.  The time at which peak nystagmus occurs varies with $1/a^2$. }
\end{center}
\end{figure}

The trajectories of three otoconia (A, B, C) falling from the cupula into the curved tapering region (Fig. \ref{fig:toroidbppv}a), as well as the associated cupular volume displacement, pressure, and nystagmus (Fig. \ref{fig:toroidbppv}b), highlight different features of BPPV attacks.  
Three distinct features can be seen in trajectory B:  1) During the first $\sim$5 seconds, the otoconium falls through the ampulla and the cupular volume displacement grows only slowly;  2) During the following $\sim$3 seconds, the otoconium falls into the narrow duct and the cupular volume displacement/nystagmus increases rapidly; and 3)  Once the otoconium approaches the wall, it no longer exerts a significant pressure on the cupula, so that the cupular volume displacement decays with time constant $\tau_c$.  The peak nystagmus for a single average-sized otoconium is approximately the same with or without velocity storage:  $\dot{E}^{{\rm max}} \approx 2^\circ$/s when velocity storage is incorporated, which is to be compared with the purely mechanical response, $\Delta V_c^{{\rm max}}\approx 30$ pL, which also gives nearly 2$^\circ$/s using Eq. (\ref{eq:maxdisp}).  Velocity storage mainly affects the time constant for nystagmus decay.

Because we have imposed a minimum particle-wall separation $d_c$, the particle moves significant distances even when close to the wall (Fig. \ref{fig:toroidbppv}a).  We interpret the initial ($\approx$ 5 seconds) period of low cupular volume displacement as `latency', and note that the delay in sensation and duration of latency predicted here are consistent with clinical observations.

The trajectories of all simulated otoconia begin just below the cupula, because this location in the posterior canal is physically the lowest for normal head orientations, and is thus assumed to be an otoconium `collection point'.  No latency should be expected for otoconia started in the duct, although a significant nystagmus can result.

\subsection{Influence of Otoconium Size upon cupular pressure and displacement}
\label{sec:sizes}

Otoconia have been found in a range of sizes, typically between 0.5-15 $\mu$m \cite{campos90}.
In this section, we provide some guidelines for the influence of otoconium size on the nystagmus and cupular volume displacement in canalithiasis.  The results of simulations for different-sized particles (Fig. \ref{fig:toroidbppv}c) demonstrate that peak nystagmus and cupular volume displacement increase linearly with particle size $a$ for particles larger than a few microns, which can be understood as follows:  Hydrodynamic interactions do little to an otoconium's trajectory until it comes very close to the wall, where the transcupular pressure is negligible.  Thus particles of different sizes follow approximately the same trajectory, but with different speeds ($U_s\propto a^2$).  Transcupular pressure is proportional to otoconium mass ($\Delta P_c\sim a^3$), but is only exerted while the otoconium is away from walls (a `settling time' $T_s\sim 1/U_s \propto a^{-2}$).  For particles large enough to settle faster than $\tau_c$ or $\tau_v$, the exponentials in the integrals of Eq. (\ref{eq:xcsol}) or Eq. (\ref{eq:nystagmussol}), respectively, are approximately constant while $\Delta P_c$ is appreciable.  In this case, peak cupular volume displacement and nystagmus ($\propto T_s \Delta P_c $) should increase linearly with particle size $a$ (Fig. \ref{fig:toroidbppv}c).

\subsection{Cupulolithiasis}
\label{sec:cupulolithiasis}

In the cupulolithiasis mechanism, otoconia are physically attached to the cupula, causing it to distort when the head is turned, so that the pressure exerted by a head suddenly turned would be constant.  The maximum cupular volume displacement and nystagmus would occur if the head were tilted and held in a position such that gravity acts normal to the cupula.  In this `worst-case' scenario, the transcupular pressure is given by
$\Delta P_c = (4 \pi /3)a^3 \Delta \rho g/\pi b_c^2$, and the corresponding cupular volume displacement is given by
\be
V_c(t) = \frac{F_g}{K \pi b_c^2}\left(1-e^{-t/\tau_c}\right),
\ee
with a nystagmus response
\be
\dot{E}(t) = \frac{\lambda_1^4 a^3\Delta \rho g}{24 \rho (\beta_u + \beta_d)R^2 b_c^2 }\tau_v \left(1-e^{-t/\tau_v}\right)\approx 0.6^\circ/s \left(1-e^{-t/\tau_v}\right).
\label{eq:cupulolithiasisn}
\ee

\section{Discussion}
\label{sec:conclusion}

Eq. (\ref{eq:pressuredrop}), expressing the transcupular pressure set up by a small particle settling through a fluid-filled canal, is perhaps the central result of our work.  It is significant that this expression follows directly from a solution of the Stokes equations for viscous flow.  While House and Honrubia's $\Delta P_H$ does capture many of the important characteristics of canalithiasis, the form of that equation was argued based on a series of assumptions (some valid, some not) rather than from a solution to the equations of motion for the fluid.  

Eq. (\ref{eq:pressuredrop}) confirms several features of House and Honrubia's expression:  (i) the characteristic magnitude of the transcupular pressure, (ii) the amplification of transcupular pressure when the particle enters a narrow section of the channel, and (iii) the inability of a force perpendicular to canal walls to result in a transcupular pressure.  However, both quantitative and qualitative corrections to $\Delta P_H$ arise, as evident from Eq. (\ref{eq:pressuredrop}):  (i) the transcupular pressure $\Delta P_c$ for a particle on the centerline is six times greater than given by $\Delta P_H$ -- a surprising consequence of viscous resistance -- and (ii) the transcupular pressure is diminished for particles located away from the canal centerline.  

Also significant is that Eq. (\ref{eq:pressuredrop}) provides a quantative estimate (subject to the assumptions of small particles far from canal walls) for the transcupular pressure in an arbitrary (slowly-varying, gently-curved) geometry, without requiring a detailed numerical analysis of the resulting fluid flow.  While we have employed a simplified model geometry, it would be straightforward to apply the present results to a more realistic, three-dimensional geometry of the semicircular canals, as well as to multi-step head maneuvers.

House and Honrubia used $\Delta P_H$ in one-dimensional simulations with otoconia constrained to fall along the canal centerline, and found that an otoconium mass $0.087 \mu$g (or 18 otoconia of radius 7.5$\mu$m) would be required to achieve a (measured) peak nystagmus of $\dot{E}\approx42^\circ$/s.  By comparison, an analogous one-dimensional simulation using Eq. (\ref{eq:pressuredrop}) would require only such three otoconia (0.014 $\mu$g) to achieve the same peak nystagmus.  In two-dimensional simulations, however, simulated particles fall away from the centerline and diminish the transcupular pressure (Fig. \ref{fig:toroidbppv}).  Therefore, approximately 20 otoconia, or $0.1 \mu$g, would be required to account for the peak nystagmus measured by House and Honrubia, since one average-sized otoconium can cause a nystagmus of approximately $2^\circ$/s.

The observed latency of BPPV can be understood as the time during which otoconia settle through the ampulla (where the transcupular pressure is small), as was also suggested by House and Honrubia.  As a result, particles that start in the duct rather than the ampulla should produce no latency \cite{house03}.

During therapeutic maneuvers, the head is re-oriented several times so that otoconia travel around the duct and into the {\sl vestibule}, a region common to all three canals (Fig. \ref{fig:defscc}a).  Two significant features can be expected:  i)  Towards the end of such a maneuver, otoconia traverse the {\sl crus commune}, where the anterior and posterior canals are joined.  During this period, the pressure field of the settling otoconia is exerted across {\sl both} the anterior and posterior canals, and the {\sl direction} of nystagmus should change correspondingly, from one that is in the plane of the posterior canal (upbeating mixed with torsion), to one in the plane of the sum of both canals (purely torsional).  ii)  As otoconia settle out of the narrow duct into wider regions (\eg the vestibule or the crus commune), the transcupular pressure is {\sl reduced} according to Eq. (\ref{eq:pressuredrop}).  This pressure reduction provides a possible resolution to Buckingham's critique of canalithiasis \cite{buckingham99}:  once in the vestibule, otoconia may still exert a fluid pressure on the cupula, the magnitude should be substantially decreased according to Eq. (\ref{eq:pressuredrop}).

Experimental investigations of BPPV models have used the vestibular systems of frogs.  Otoconia were inserted into the open end of an isolated posterior semicircular canal (PSC) \cite{suzuki95}, or dislodged from the utricle and driven into the PSC of an intact labyrinth \cite{otsuka03}.  Neural activity was measured following a change in PSC orientation, with free-floating otoconia (`canalithiasis') and attached otoconia (`cupulolithiasis').  For quanitative comparisons with the present theory, several ingredients would be required:  physical parameters of the frog PSC, the relation between firing rate of PSC nerves and rotational motion, and the mass of otoconia in the experiment.  Nevertheless, qualitative features agree with our model.  In `cupulolithiasis' experiments, little or no latency was evident, and the change in firing rate was sustained, with decremental time constants of 16.8$\pm$4.9 s and $>$43 s.  In `canalithiasis' experiments, latency periods of 2-4 s were obserived, and decay time constants of 7-10 seconds were measured.  

The experimental results \cite{suzuki95} contrast with the predictions of the present theory, the theory of House and Honrubia, and clinical observations \cite{house03, semont88} in one significant aspect:  a latent period was measured for a clump of otoconia started in the duct, which Suzuki \etal attributed to a re-orientation of the clump itself.  It should also be noted that in these experiments, otoconia typically travelled in `clumps' and slid along side walls.

The effects of particle inertia were estimated and demonstrated to be negligible from a hydrodynamic standpoint, which seems to run counter to the philosophy of clinical maneuvers designed with particle inertia in mind.  Although inertia is irrelevant in the trajectories of small settling particles, it may play a significant role in dislodging otoconia adherent to canal walls.

A large range of eye movement responses are possible for the various sizes, trajectories, and collective behavior of otoconia.  Multiple otoconia may be clumped or dispersed, and otoconia clumps may be treated by our methods so long as the clump is smaller than the duct radius.  For example, a clumped group of $N$ otoconia would behave like a single larger particle (of radius $a' \approx N^{1/3} a$), giving a maximum cupular volume displacement which is approximately $N^{1/3}$ times larger than for a single otoconium.  On the other hand, a group of $N$ dispersed otoconia settling independently would result in approximately $N$ times greater cupular volume displacement and nystagmus than for a single otoconium.   

Clumped otoconia thus cause less nystagmus than dispersed ones, a prediction that is seemingly at odds with the fatiguability of BPPV.  Peak nystagmus typically decreases with repeated head maneuvers \cite{brandt91}, a fact which is typically attributed to the breaking and dispersion of particle clumps and/or particle margination \cite{parnes93}.  
Several mechanisms might reconcile these pictures:  i)  Particle margination:  repeated maneuvers presumably break up clumps and disperse the particles, which when isolated might adhere to the membraneous canal walls. ii) Exceptionally large particle clumps that occupy almost all of the duct would exert a larger transcupular pressure, and would move significantly more slowly, than individual particles.  This would result in greater nystagmus, which would be lessened by the breakup of such a large clump.  iii) The response of dispersed particles could result in weaker nystagmus than a single clumped stone if the dispersed particles follow different trajectories, since the nystagmus response is sensitive to particle trajectory (Fig. \ref{fig:toroidbppv}). 

A comparison of the cupular volume displacement and nystagmus due to a single otoconium in both canalithiasis and cupulolithiasis suggests that canalithiasis is the stronger mechanism, and that multiple and/or larger particles are necessary to produce the same response in cupulolithiasis than in canalithiasis.  While counter-intuitive, this follows from Eq. (\ref{eq:pressuredrop}) for canalithiasis, which provides for a pressure amplification by a factor of up to 2$b_c^2/b_d^2 \approx 36$ over cupulolithiasis.  In cupulolithiasis, if the head were tilted and held in place indefinitely, an average otoconium attached to the cupula would yield a cupular volume displacement $V_c \approx 2$ pL and nystagmus response $\dot{E}\approx 0.6^\circ/$s, smaller than the corresponding canalithiasis results by factors of 10 and 3-4, respectively.  Furthermore, the nystagmus due to a single, average otoconium in cupulithiasis is below the sensation threshold, meaning that multiple and/or larger otoconia are required for cupulolithiasis.  


\appendix
    
\section{An application of the Reciprocal Theorem  
for a point force in a tube of slowly varying radius}  
\label{app:recip}

In this appendix we use the reciprocal theorem from
low-Reynolds-number hydrodynamics \cite{leal92} to determine the pressure drop at
large distances from a point force (Stokeslet) in a tube. This
analysis was apparently first given by Brenner \cite{brenner58} for a point force in a
circular cylinder and we show that the same ideas can be extended naturally to a tube of arbitrarily varying cross section and arbitrary orientation relative to the direction of gravity.  

Recall that the reciprocal theorem relates any two solutions $(\mathbf{u}, \mathbf{\sigma}, \mathbf{f})$ and $(\hat{
\mathbf{u}},\hat{\mathbf{\sigma}}, \hat{\mathbf{f}})$ to the Stokes
equations 
\bea
\nabla \cdot {\bf u} &=&0\\
\nabla \cdot {\mathbf \sigma} + {\bf f} &=&{\bf 0},\label{eq:stokeseq}
\eea
where $\mathbf{\sigma}$ is the stress tensor (with components $\sigma_{ij}$),
\be
\sigma_{ij}=-p\delta_{ij}+\mu\left(\pd{u_i}{x_j}+\pd{u_j}{x_i}\right).
\ee 
Here $\mathbf{u}$ and $\mathbf{f}$ represent,
respectively, the velocity field and force per unit volume
acting on the fluid.  Then, within a volume $V$ and
corresponding bounding surface $S$, these fields are related by
\begin{equation}  
\int_{S}{\mathbf{n \cdot \sigma \cdot \hat{u}}  ~{\rm d}S} 
- \int_{S}{\mathbf{n   
\cdot \hat{\sigma} \cdot u}  ~{\rm d}S}
 = -\int_{V}{\mathbf{u \cdot \hat{f}} ~{\rm d}V}   
+\int_{V}{\mathbf{\hat{u} \cdot f} ~{\rm d}V}~,   
\label{recip-thrm1}
\end{equation}  
where ${\bf n}$ is the unit normal directed into the fluid domain from the surface.

Here we consider the specific case where $(\mathbf{u}, \mathbf{\sigma}, \mathbf{f})$ represent the
fields associated with a Stokeslet in an otherwise quiescent fluid inside an infinitely long tube of
arbitrary cross-sectional shape, and where $(\hat{
\mathbf{u}},\hat{\mathbf{\sigma}}, \hat{\mathbf{f}})$ represents the steady pressure-driven
flow in a non-uniform duct.  In this case, the
volume $V$ corresponds to the interior of the tube.  Now,  due to the no-slip condition on
the walls of the duct, the surface integrals in equation
(\ref{recip-thrm1}) are only nonzero at the `ends' of the duct, which
we consider to be at a very large distance (many cylinder radii) from the
point force, where the Stokeslet velocity field $\hat{\mathbf{u}}$ is negligibly small.
  
We now simplify (\ref{recip-thrm1}) in steps.  First, the  
pressure-driven flow contains no body forces in $V$, ${\hat{\mathbf{f}}} =
{\bf 0}$.  Second, the flow due to a Stokeslet in a bounded cylindrical domain 
decays exponentially with axial distance from the singularity \cite{blake79}, 
so $\mathbf{u} \rightarrow \mathbf{0}$ towards the ends of the duct.
Further, since $\mathbf{f}$ corresponds to a Stokeslet of strength ${\bf F}$, then
$\mathbf{f}={\bf F} \delta ({\bf r})$, where $\delta({\bf r})$ denotes the Dirac delta function, and
we may evaluate the second integral on the right-hand side of (\ref{recip-thrm1})
as $\mathbf{\hat{u}}_p(\mathbf{x}_p)\mathbf{\cdot F}$, where  
$\mathbf{\hat{u}}_p(\mathbf{x}_p)$ is the pressure-driven velocity field
 at the  location ${\bf x}_p$ of the Stokeslet.  
  
Since the `ends' of the cylinder are considered far from the Stokeslet  
where ${\bf u}\rightarrow {\bf 0}$, then the corresponding stress field  
is simply a pressure that remains constant across  
each `end' section.  If we denote the difference in pressures between  
the two `ends' as $\Delta P$, then the first term
on the left-hand side of (1) reduces to  
$\hat{Q} \Delta P$, where $\hat{Q}$ is the volume flux due to the  
pressure-driven flow.  Combining the above results gives   
\begin{equation}  
\Delta P = \frac{\bF \cdot {\bf \hat{u}}_p}{\hat{Q}}  ~.
\label{eq:recippres}
\end{equation} 
This holds for a tube of arbitrarily varying cross-section and and arbitrarily-oriented force ${\bf F}$.

Finally, we invoke the lubrication approximation, 
which provides that for a circular tube of slowly varying 
radius $b(s)$ the flow velocity is given by a 
generalized Poiseuille flow,
\be
{\bf \hat{u}}= U_0(s) \left(1 - \frac{r^2}{b(s)^2}\right)\zh,
\ee
where $s$ is the coordinate along the cylinder axis, and $r$ is the radial coordinate, measured from the centerline. 
Using the corresponding flow rate,
\be
\hat{Q} = \frac{\pi b(s)^2U_0(s) }{2},
\ee
in Eq. (\ref{eq:recippres}), we find the Stokeslet sets up a pressure difference
\begin{equation}  
\Delta P = \frac{2\bF \cdot \zh}{\pi b(s_p)^2} \left(1-\frac{r_p^2}{b(s_p)^2}\right).
\label{eq:qed}
\end{equation} 

Note that the exact location of the `ends' does not matter, so long
as the Stokeslet flow (${\bf u}$) is negligibly small at the ends.  Eq. (\ref{eq:pressuredrop}) follows from Eq. (\ref{eq:qed}) by taking the ampullar and vestibular sides of the cupula as the `ends'.  This implies, however, that
Eq. (\ref{eq:pressuredrop}) does not hold for particles `close' (within a few canal radii) to the cupula,
since in that case ${\bf u}$ differs from the assumed `infinite tube' solution.  For example, a Stokeslet located much closer to the cupula than to the side walls exerts an average transcupular pressure $\bF/\pi b_c^2$ \cite{blake71}, which is half as large as that predicted by \ref{eq:pressuredrop}.

Lastly, we examine the error introduced by the point-force approximation.  Because variations in the canal radius are gradual, an error estimate can be obtained from results 
The pressure drop due to a sphere settling through an otherwise quiescent fluid in an infinitely-long, straight, circular cylinder, is given by Eq. (\ref{eq:qed}), with errors of order 
$\oo{\left(a/b)^2}\right)$, 
when the sphere is not within a few radii of the wall (see Eqs. 7-3.96-7.3.98 in \cite{happel}).  This corresponds to less than one percent for an average otoconium.  Corrections due to canal curvature of order $b_d/R$, approximately 5\%, may also be expected.

\section{Cupular response to a particle settling in a straight, tapering channel}
\label{sec:bppvstraight}

\renewcommand{\theequation}{B\arabic{equation}}
\renewcommand{\thefigure}{B\arabic{figure}}

A straight, tapering channel provides a particulary simple illustration of
the pressure amplification provided by a narrowing channel, which is expected to be a significant dynamical feature
accompanying sedimentation of an otoconium, and highlights the hydrodynamic explanation for latency.

\begin{figure}
\begin{center}
\includegraphics[height=3in]{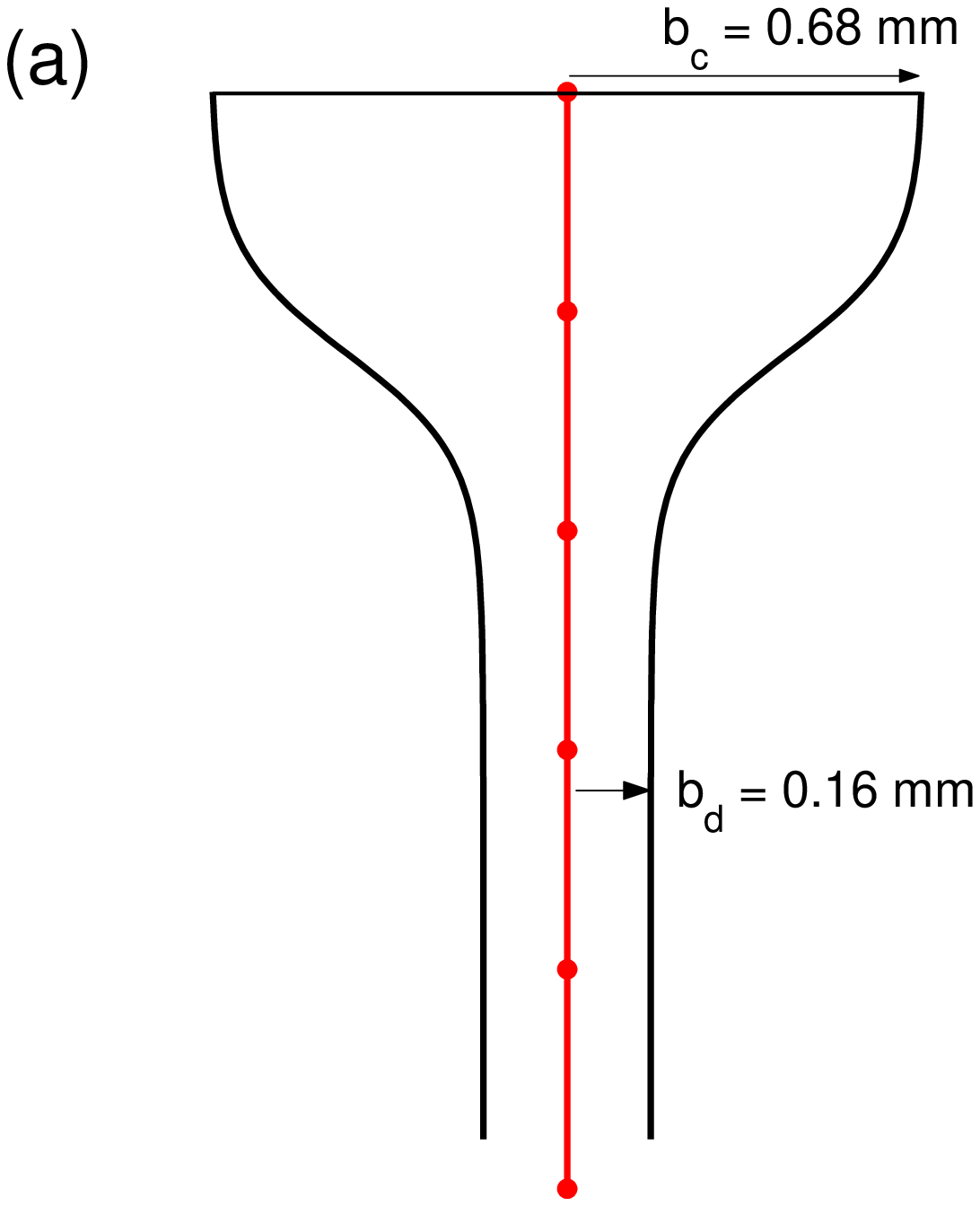}\hfill
\includegraphics[height=3in]{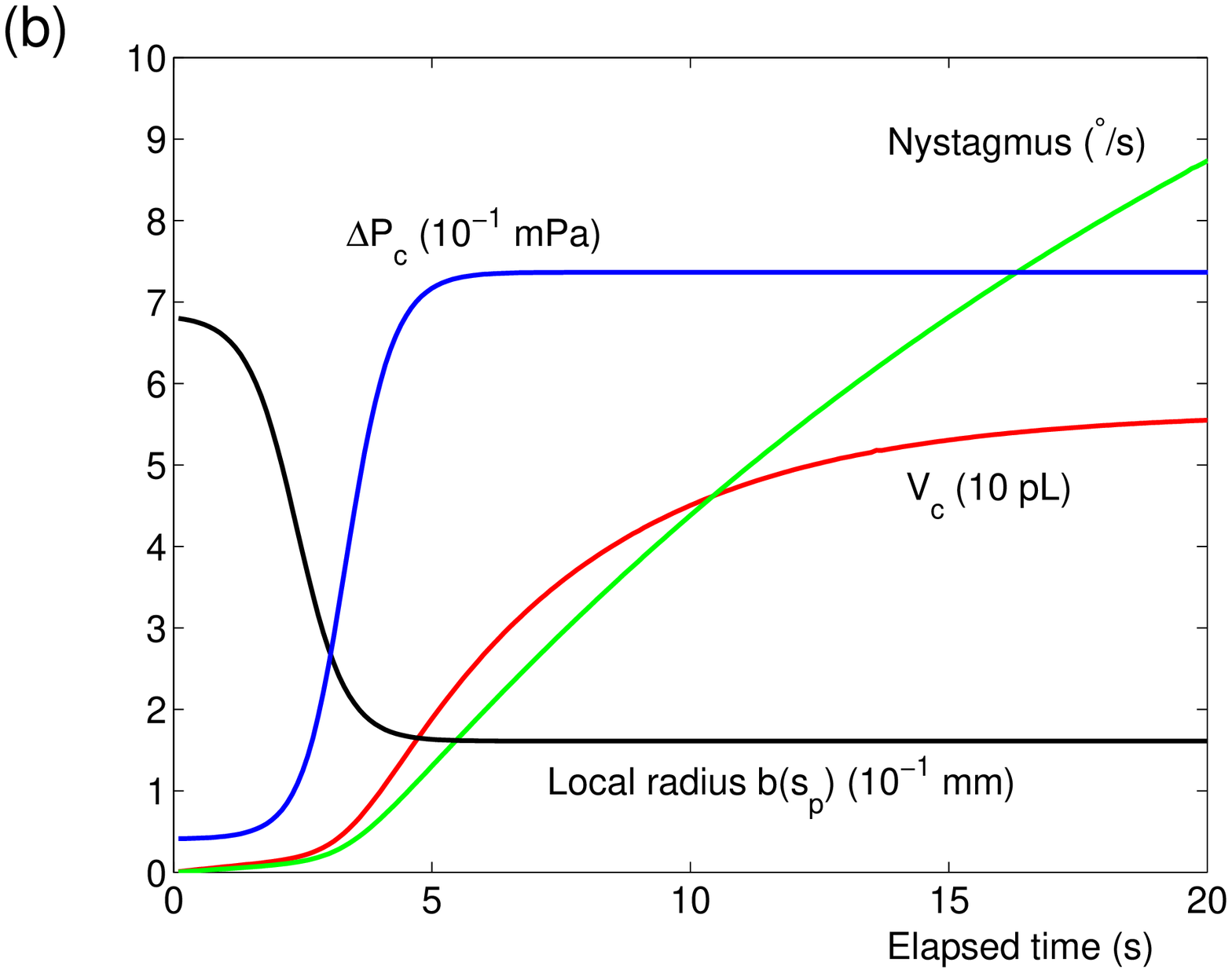}
\caption{\label{fig:straighttaper} (a) Trajectory of an otoconium falling through the straight, tapering
canal described by Eq. (\ref{eq:straighttaper}).  Heavy, filled dots denote the otoconium position every
two seconds.  (b)  Cupular pressure $\Delta P_c(t)$, displacement $V_c(t)$ and nystagmus 
$\dot{E}(t)$ as an otoconium falls along the centerline of the straight, tapering channel (whose profile is also indicated).  The settling otoconium sets up a pressure field that is small when 
the otoconium is in the wide part of the channel ($t < 3$ seconds) and jumps by a 
factor $b_c^2/b_d^2 \approx 18$ when the otoconium enters the narrow part of the channel.
Correspondingly, the cupular volume displacement increases slowly while the force is small 
($\approx 3$ seconds), then increases over a timescale $\tau_c$ to the large, constant force
as the otoconium falls through the duct.  The nystagmus response is similar, but increases over
the longer time scale $\tau_v$.  Several seconds of `latency' can be seen as the particle settles through
the wide part of the channel, after which the force-amplifying effect of the narrowing canal sets in.  Geometrical and physical parameters are given in Table I.}
\end{center}
\end{figure}

We consider a particle of radius $a$ that settles along the midline of a straight
vertical channel of varying radius $b(s)$
\be
b(s) = \frac{b_c + b_d}{2} -\frac{(b_c - b_d)}{2} \tanh\left(\frac{s-s_0}{R \delta}\right),
\label{eq:straighttaper}
\ee
so that $b(-\infty)= b_c$ and $b(\infty)= b_d$ (Fig. \ref{fig:straighttaper}a).  
The specific functional form in Eq. (\ref{eq:straighttaper})
has been chosen for convenience but has the feature of narrowing from $b_c$ to $b_d$ over a distance
$R\delta$, which captures one significant geometric feature of the semicircular canal (Fig. 1b).
Neglecting hydrodynamic interactions with the wall, the sphere settles at the Stokes settling velocity,
so that its axial position is given by $s_p(t) = U_s t$, where $U_s$ is given by Eq. (10).
Viscous resistance to the settling sphere sets up a time-dependent pressure drop (Fig. \ref{fig:straighttaper}b) on the cupula found by combining Eqs. (8), and (\ref{eq:straighttaper}),
\be
\Delta P_c(t) = \frac{32 a^3 \Delta \rho g}{3 \left\{(b_c + b_d) - (b_c - b_d) \tanh[( U_s t-s_0)/R \delta]\right\} ^2}.
\ee

As the particle settles through the narrowing portion of the channel, the pressure it exerts on the cupula increases by a factor $b_c^2/b_d^2\approx 18$, and asymptotically approaches a constant value $\Delta P_c^\infty= 8 a^3 \Delta \rho g/3 b_d^2.$
The cupular volume displacement (Fig. \ref{fig:straighttaper}b) is then given by Eq. (7).  After many time constants $\tau_c$, the cupular volume displacement asymptotically approaches its limiting value
\be
V_c^\infty = \frac{\Delta P_c^\infty \tau_c}{\gamma} = \frac{8 a^3 \Delta \rho g}{3 K b_d^2} \approx 57{\rm pL},
\ee
where we have evaluated $V_c^\infty$ using the typical values reported in Table I.
The nystagmus $\dot{E}(t)$ (Fig. \ref{fig:straighttaper}b), calculated using Eq. (14), responds over the longer time scale $\tau_v$, and asymptotically approaches its limiting value,
\be
\dot{E}^\infty = \frac{\lambda_1^4 \Delta P_c^\infty \tau_v}{32 \rho (\beta_u + \beta_d) R^2} \approx 0.28\,{\rm rad,}\,\,{\rm or}\,\,16^\circ.
\ee
This simulation is representative of basic hydrodynamic processes and corresponding cupular responses, but is nevertheless unrealistic, as the real semicircular canal is curved and the forcing is 
truncated when the otoconium hits the wall.  It does, however, clearly demonstrate the pressure-amplification
effect of the tapering canal. 


\section{Summary of results for a sphere settling in a circular cylinder}
\label{app:hydrocoupling}
\renewcommand{\theequation}{C\arabic{equation}}
\renewcommand{\thefigure}{C\arabic{figure}}

This appendix summarizes several results for the hydrodynamic influence of the
cylinder walls upon the motion of a small sedimenting particle.  Our motivation
for including this appendix is twofold:  1)  to explain how hydrodynamic 
coupling between the wall and the particle can be treated in a reasonably 
straightforward yet accurate manner, and 2) to demonstrate
that this hydrodynamic coupling is largely unimportant in modelling 
the sedimentation of small particles in the semicircular canals.
In regard to the second point, the walls only exert a significant influence
upon the motion of a sedimenting particle when the particle is very close to
the wall. In this limit, however, the fluid pressure on the cupula due to the
sedimenting particle (given by Eq. (8)) is exceedingly small.  
Therefore, wall-particle interactions are typically not significant for modelling 
cupular volume displacement, but are important when the trajectories of the particles are themselves of interest.

A simple expression for the sedimentation velocity of a sphere in a geometry
as complicated as that of the semicircular canals is unavailable.  However,
because the radius of curvature $R$ of the torus is much larger than the 
canal radius $b$, we approximate the canals as straight
circular cylinders, which involves errors of order $\oo{b/R}$.  Even for this
comparatively simple geometry, no uniformly valid expression is known for the motion 
of a sphere through a viscous fluid in a circular cylindrical container.  
Rather, numerical and asymptotic formulae have been derived in various limits.  In this appendix,
we summarize several results, which we use in our simulation of canalithiasis.

We consider a small spherical particle of radius $a$, with externally applied 
force $\bF=F_\| {\bf \hat{z}} + F_\perp {\bf \hat{r}}$,
located at radius $r_p$ in a circular cylinder of radius $b$.  We introduce two dimensionless parameters, $\beta = r_p/b$ and $\kappa = a/b$ 
to characterize the system.  The parameter $\beta$ indicates the dimensionless radial position of the particle, and $\kappa$ gives a dimensionless particle radius.

In modelling the sedimentation of otoconia in the semicircular canals, we are concerned 
with the motion of particles that are small 
with respect to the size of the cylinder, and thus restrict our attention to
the regime $\kappa \ll 1$.  Within this limit, there are two primary regimes:
1)  $1-\beta \gg \kappa$, where the distance from the particle to the wall is 
large compared to the particle size, and 2)  $1 - \beta \sim \kappa$, where
the particle is close to the wall.

\begin{figure}
\begin{center}
\includegraphics[width=3in]{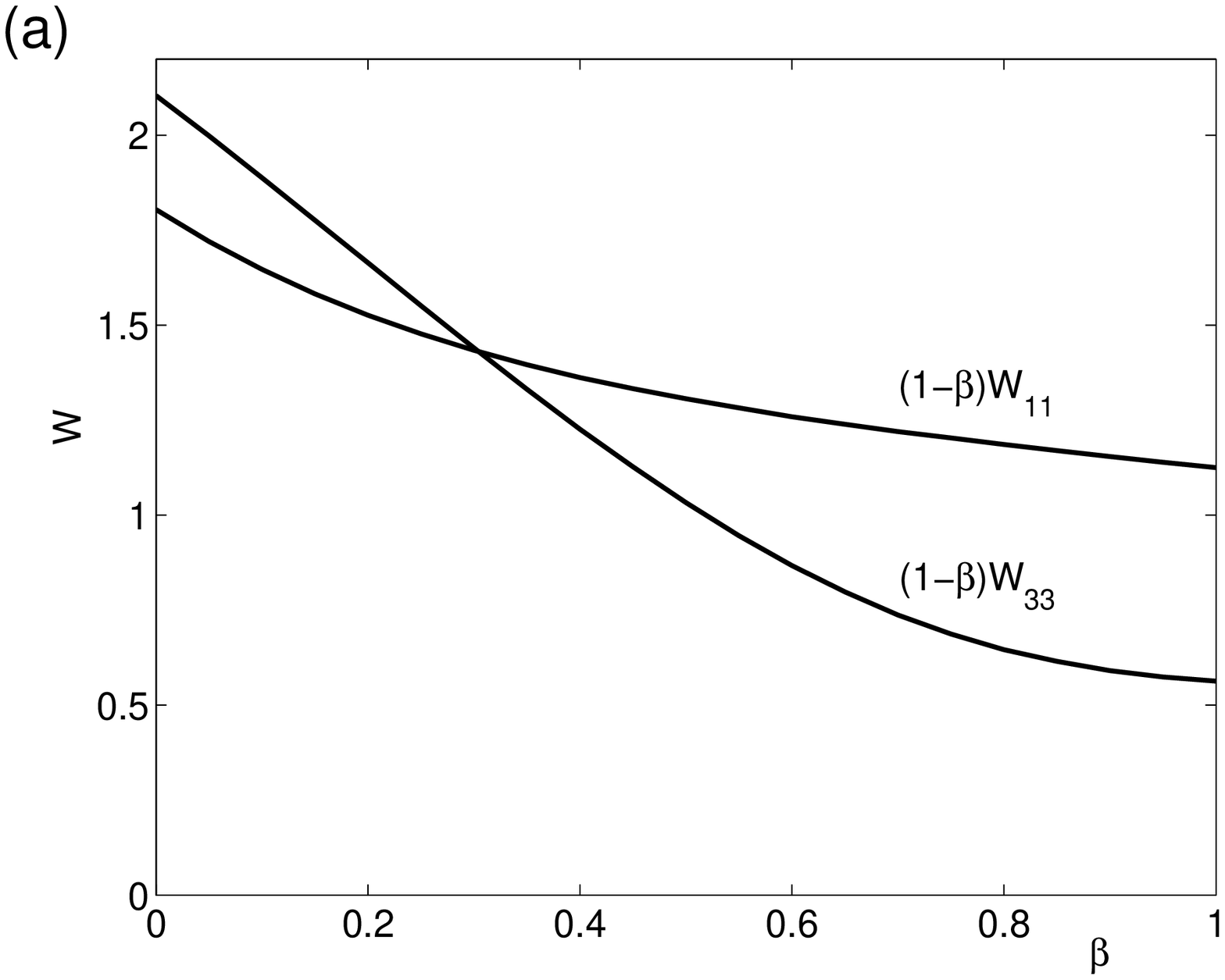}\hfill
\includegraphics[width=3in]{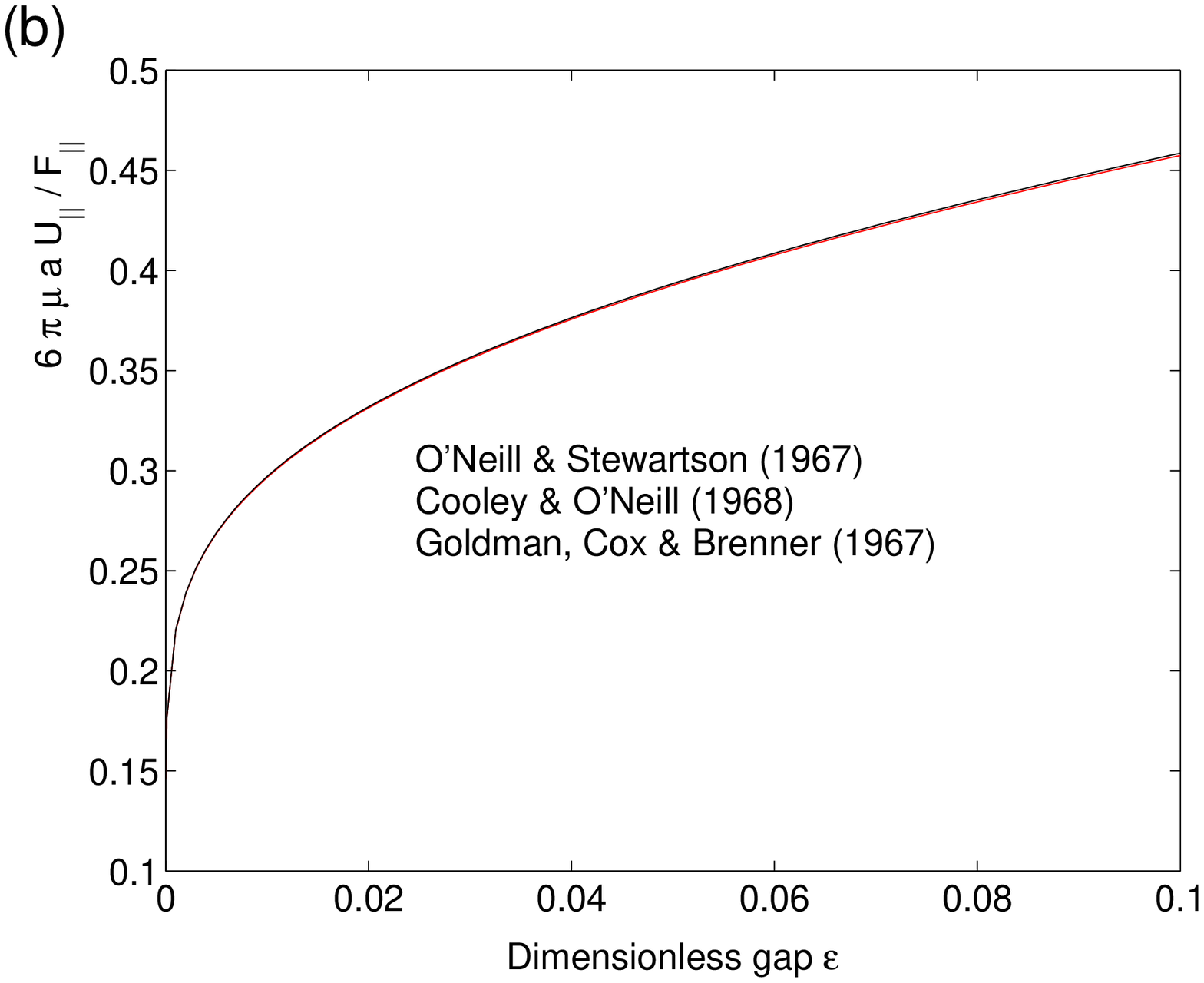}
\caption{\label{fig:hirschws} (a)  Numerical factors for Eqs. (\ref{eq:uperp}) giving corrections to the velocity of a sphere in an infinite, straight, circular cylinder which is forced radially ($W_{11}$) or axially ($W_{33}$).  Here $\beta = r_p/b$ is a measure of the sphere's fractional distance to the wall.  Both $W_{11}$ and $W_{33}$ diverge like $(1-\beta)^{-1}$ as the sphere nears the wall, which is the reason we have plotted $(1-\beta)W_{11}$ and $(1-\beta)W_{33}$.  (From (Hirschfeld {\sl et al.}, 1984), Table 2.)  (b) Settling velocity of a sphere parallel to the wall, in the limit where the sphere-wall gap is much smaller than the sphere radius (Eqs. \ref{eq:goldman} and \ref{eq:oneill}).}
\end{center}
\end{figure}

For the first regime, a method of reflections can be used to perturbatively calculate the 
influence of the wall upon the motion of the sphere. 
While results accurate to $\oo{\kappa^2}$ are available (Hirschfeld {\sl et al.}, 1984), we retain terms to $\oo{\kappa}$ only, with perpendicular and parallel velocities are given by
\be
U_{\perp}(r_p \gg a) = \frac{F_\perp}{6 \pi \mu a}\left(1 - \kappa W_{11}\right)\label{eq:uperp}\,\,\,\,\mbox{and}\,\,\,\,
U_{||}(r_p \gg a) = \frac{F_\|}{6 \pi \mu a}\left(1 - \kappa W_{33}\right)\label{eq:upar},
\ee
where $W_{11}$ and $W_{33}$ are dimensionless functions (Fig. \ref{fig:hirschws}). 

A different behavior is found when the sphere is close to the wall, and the dimensionless gap $\ep=(b-r_p)/a$ is 
small. The leading-order approximation to the settling speed of the sphere can be found by treating the cylinder wall as locally planar, in which limit asymptotic formulae have been determined using a lubrication approach (Cooley \& O'Neill, 1968; Goldman \etal, 1967; O'Neill \& Stewartson, 1967).  The formula given by Goldman \etal contains an error, and should read 
\be
U^G_{||} \approx\frac{F_{||}}{6 \pi \mu a} \left[\frac{1.91-2\ln(\ep)}{1.59-3.19\ln(\ep)+\{\ln(\ep)\}^2}\right],
\label{eq:goldman}
\ee
which very nearly agrees with the results obtained in (Cooley \& O'Neill, 1968; O'Neill \& Stewartson, 1967), which give
\be
U_\|^O \approx\frac{F_{||}}{6 \pi \mu a} \left[\frac{1.85-2\ln(\ep)}{1.52 - 3.15\ln(\ep)+\{\ln(\ep)\}^2}\right].
\label{eq:oneill}
\ee
The velocity parallel to the wall thus decays logarithmically slowly.


Motion perpendicular to a planar wall due to a force $F_\perp$ can be found using the lubrication approximation (\eg Batchelor, 1967, p. 228, problem 1) in the limit $\ep\ll 1$, giving
\be
U_\perp = \frac{F_\perp \ep}{6 \pi \mu a},
\ee
so that a sphere exponentially approaches the wall with time constant $\tau_\perp = 6 \pi \mu a^2/F_\perp$.

\end{document}